\documentclass[11pt]{article} \usepackage{mystyle-new}
\usepackage{epsfig,amsmath} \usepackage{graphics}
\usepackage{cite,./mcite}
\usepackage{epsfig,amsmath} 
\usepackage{hepnames,hepunits}
 
\usepackage{authblk}

                \def\lsim{\mathrel{\rlap{\lower4pt\hbox{\hskip1pt$\sim$}}
    \raise1pt\hbox{$<$}}}                \def\gsim{\mathrel{\rlap{\lower4pt\hbox{\hskip1pt$\sim$}}
    \raise1pt\hbox{$>$}}}

\newcommand{\pt}{\ensuremath{p_{\rm T}}}

\def\cascade{{\sc Cascade3}}

\def\powheg{{\sc Powheg}}
\def\powhegbox{{\sc Powheg-BOX}}

\def\mcatnlo{{\sc mc@nlo}}

\def\hdamp{{\texttt{hdamp}}}

\usepackage{lineno}
\makeindex
\begin{document}

\title{Drell Yan production in the parton branching method and \powheg\ matching}

\author[3]{A.~Bermudez~Martinez} 
  \affil[1]{Deutsches Elektronen-Synchrotron DESY, Germany}
\author[1]{L.I.~Estevez~Banos}
\author[2]{M.~Fern\'andez~Moreira}
  \affil[2]{Higher Institute of Technology and Applied Sciences, University of Havana, InSTEC}
\author[2]{J.~Girones~Dominguez}
\author[1]{H.~Jung}
\author[2]{R.~Ramos~Blazquez}

\begin{titlepage}
\maketitle
\thispagestyle{empty}
\vspace*{-9.5cm}
\begin{flushright}
DESY-21-221 \\
\end{flushright}
\end{titlepage}

\begin{abstract}
As part of the DESY summer student program 2021, transverse momentum dependent (TMD) parton distributions obtained from the Parton Branching (PB) method were combined with next-to-leading-order (NLO) using the \powheg\ method. Computations of the resulting Drell-Yan (DY) transverse momentum spectrum were performed. A good agreement of the theoretical predictions with the measurement performed by the CMS experiment at the center-of-mass energy of 13 TeV is found, at low and intermediate DY \pt. The new scale choice option for the matching has been included in the \cascade\ event generator.
\end{abstract}

\section{Introduction} 
\label{Intro}
The precise description of the Drell-Yan~\cite{Drell:1970wh} (DY) \pt\ spectrum in $\Pp\Pp$ collisions requires higher-order perturbative calculations at large transverse momentum \pt  and the resummation of soft gluons at all orders at small \pt ~\cite{Dokshitzer:1978yd,Parisi:1979se,Curci:1979am,Altarelli:1984kp,Collins:1984kg} as it is done in analytical resummation~\cite{Bizon:2018foh,Bizon:2019zgf,Catani:2015vma,Scimemi:2017etj,Bacchetta:2019tcu,Bacchetta:2018lna,Ladinsky:1993zn,Balazs:1997xd,Landry:2002ix,Nadolsky,Alioli:2015toa,Bozzi:2019vnl,Baranov:2014ewa}, and collinear parton shower approaches~\cite{Sjostrand:2014zea,Bellm:2015jjp,Bahr:2008pv,Gleisberg:2008ta} matched to higher-order calculations~\cite{Frixione:2003ei,Frixione:2002ik,Frixione:2007vw,Nason:2012pr,Alwall:2014hca,Frederix:2015eii}. In ~\cite{BermudezMartinez:2019anj,BermudezMartinez:2020tys} the proposed combination of the Parton Branching formulation~\cite{Hautmann:2017xtx,Hautmann:2017fcj} of TMD evolution with next-to-leading order (NLO) calculations using the \mcatnlo\ method provided a good description of the DY \pt\ spectrum as measured by the CMS~\cite{CMS:2019raw}, ATLAS~\cite{ATLAS:2015iiu}, NuSea~\cite{NuSea:2003qoe,Webb:2003bj}, R209~\cite{Moreno:1990sf}, and PHENIX~\cite{PHENIX:2018dwt} experiments, corresponding to a wide range of DY masses and center-of-mass energies. The sensitivity of the predictions to non-perturbative contribution from the TMD was found small. In addition, in~\cite{Martinez:2021chk} a TMD multi-jet merging approach at leading-order was developed which provides a very good description over the whole DY \pt\ spectrum.

An alternative NLO matching method is \powheg\ ~\cite{Frixione:2007vw}. The \powheg\ formulation produces positive weighted events, and as opposed to \mcatnlo, the parton level events it provides do not depend on the specific parton shower used to simulate the parton radiation. In this paper we apply the PB approach to calculate the DY \pt\ spectrum at NLO using the \powheg\ method. We use TMDs obtained from a fit to inclusive deep-inelastic scattering (DIS) precision data~\cite{BermudezMartinez:2018fsv} from HERA. This work was done as part of the DESY summer student program 2021 were young scientists connected regularly from remote, working in a virtual environment to achieve the results hereby presented.

\subsection{Matching PB-TMDs with \powheg\ NLO calculations}

In the following we will describe how to use the PB-TMDs together with higher order perturbative calculations using the \powheg\ method. We make use of the \powhegbox\ ~\cite{Frixione:2007vw,Alioli:2010xd,Nason:2004rx} (version 2). \powheg\ generates one emission similar to the first step of a parton shower, with the difference that the evolution kernel is given by the ratio of the real emission and Born matrix element, rather than the splitting functions. Events generated by \powheg\ are weighted with a Sudakov factor determined by the exponentiation of the \powheg\ evolution kernel. 

The NLO generator \powhegbox\ produces events in the LHE format~\cite{Alwall:2006yp} which can then be read by parton shower event generators. We use the events output by \powhegbox\ and apply the PB-TMDs to modify the kinematics of the initial-state partons (and correspondingly the final state) according to the transverse momentum distributions given by the PB-TMDs. The addition of the transverse momentum to the system leads to shifts in the longitudinal momentum of the incoming partons while the invariant mass and rapidity of the partonic system are required to remain unchanged.

\begin{figure}
\begin{center} 
\includegraphics[width=0.45\textwidth]{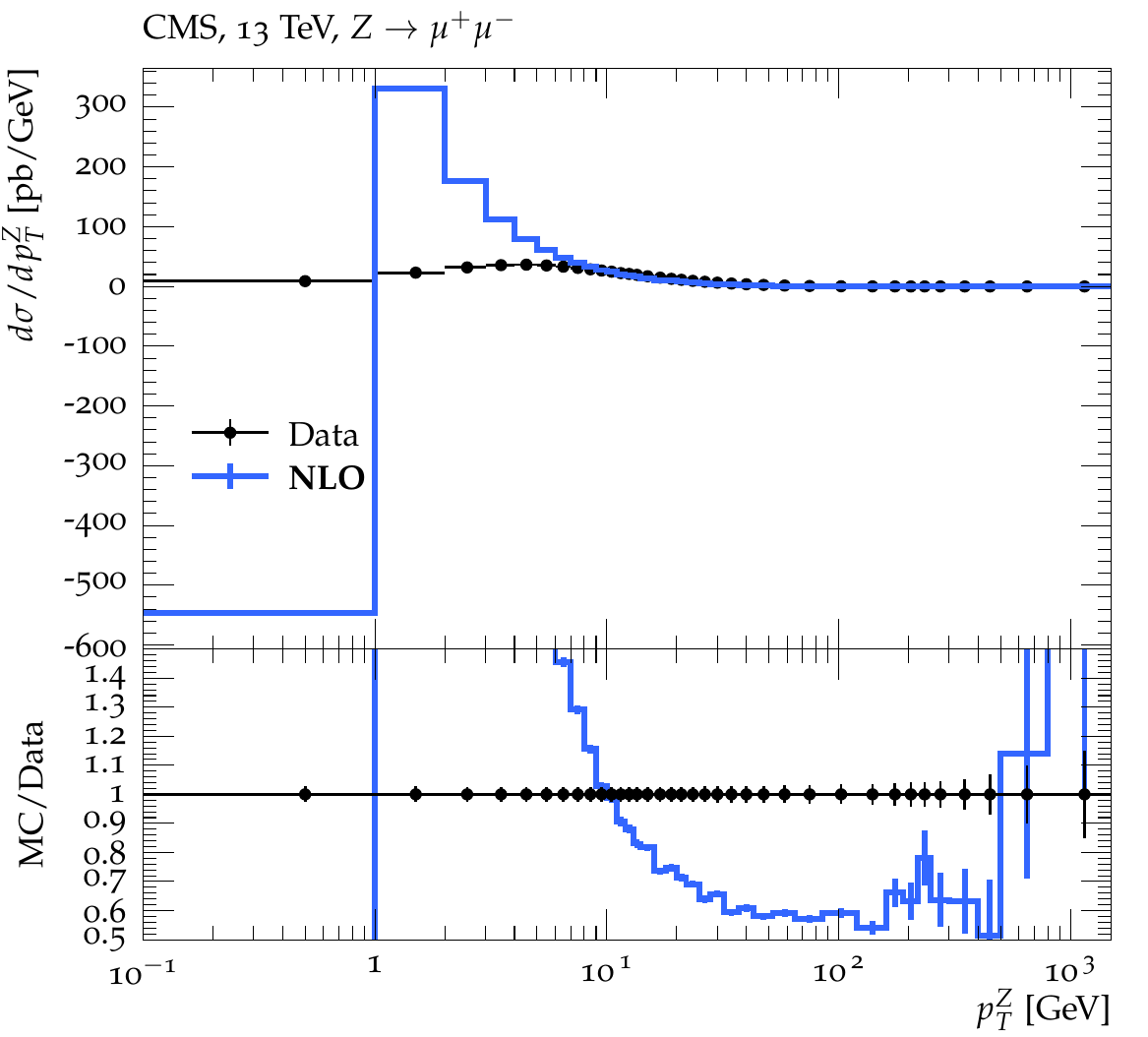}
\includegraphics[width=0.45\textwidth]{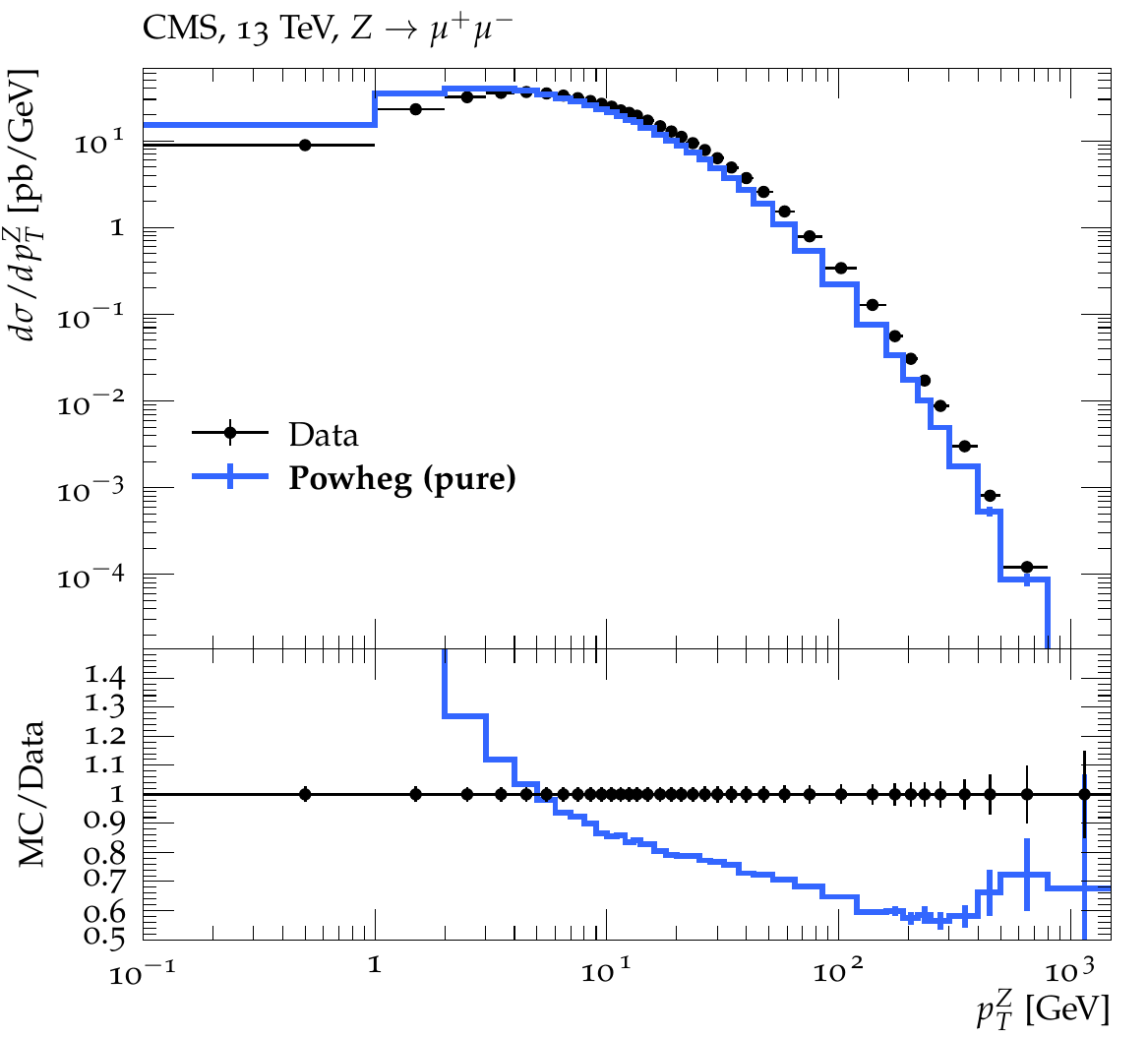}
  \caption{\small  (Left) DY \pt\ spectrum as obtained from a pure NLO calculation. (Right) DY \pt\ spectrum as obtained from a pure \powheg\ (LHE level). The calculations are shown together with the measurement of the DY \pt\ spectrum by the CMS experiment at the center-of-mass energy of 13 TeV~\protect\cite{CMS:2019raw}.}
\label{fig1}
\end{center}
\end{figure}

The DY transverse momentum spectrum calculated at pure NLO within the \powhegbox\ is shown in Fig.~\ref{fig1} (left). In order to obtained the pure NLO prediction we suppressed the action of the \powheg\ Sudakov by chosing a very small value for the \hdamp\ parameter ($\hdamp=0.001$). In Fig.~\ref{fig1} (right) the pure \powheg\ calculation (LHE level) without any suppresion of the \powheg\ Sudakov is shown. The predictions are overlaid with the DY  \pt\ spectrum measurement performed by the CMS collaboration at 13 TeV of center-of-mass energy~\cite{CMS:2019raw}. 

As observed in Fig.~\ref{fig1}, the \powheg\ method provides a positive spectrum at the LHE level compared to pure NLO, as well as compared to \mcatnlo ~\cite{BermudezMartinez:2019anj}. The potential of a combined PB-TMD + \powheg\ calculation can be seen in Fig.~\ref{fig1} (right), where a poor descrition by pure \powheg\ of low and intermediate DY \pt\ is observed.

\begin{figure}
\begin{center} 
\includegraphics[width=0.45\textwidth]{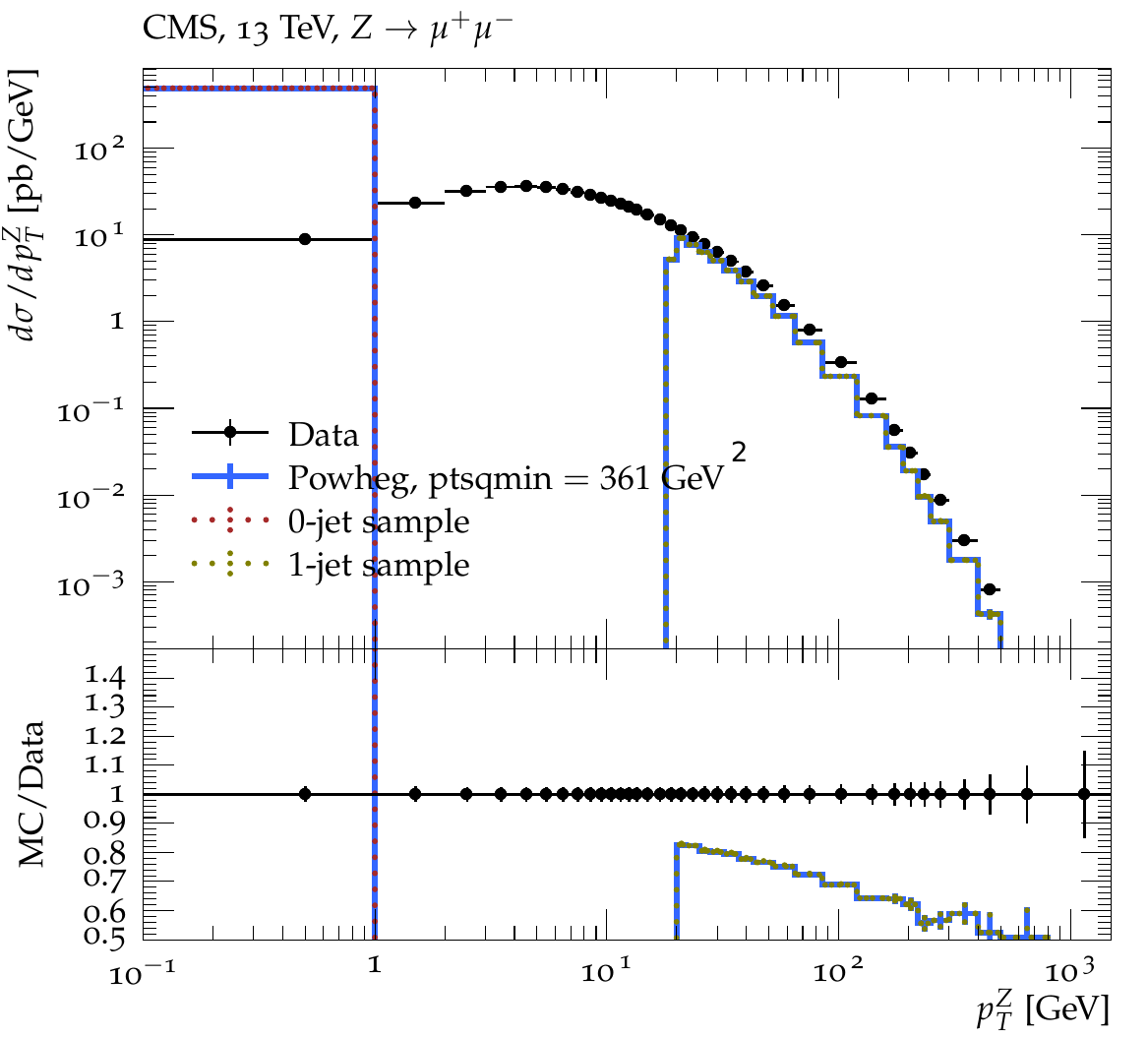}
  \caption{\small DY \pt\ spectrum as obtained from a pure \powheg\ (LHE level) setting $\texttt{ptsqmin} = \text{361 GeV}^2$. The calculation is shown together with the measurement of the DY \pt\ spectrum by the CMS experiment at the center-of-mass energy of 13 TeV~\protect\cite{CMS:2019raw}.}
\label{fig2}
\end{center}
\end{figure}

In order to match the PB-TMDs to the \powheg\ calculation we set the $p_{\perp,\text{min}}$ parameter in \powheg\ by changing the input variable \texttt{ptsqmin}. Below $p_{\perp,\text{min}}$ \powheg\ produces n--parton kinematics while above $p_{\perp,\text{min}}$ it produces (n+1)--parton events. In Fig.~\ref{fig2} we show the result at the LHE level of the \powheg\ calculation using $\texttt{ptsqmin} = \text{361 GeV}^2$.

In order to avoid double counting between the real emission produced by \powheg\ and emissions from the PB-TMD parton shower the matching scale is set equal to the parameter \texttt{SCALUP} (included in the LHE \powheg\ output). The \powhegbox\ generator by default attributes a value $\texttt{SCALUP} = \sqrt{\texttt{ptsqmin}}$ which is then used as the upper scale for the transverse momentum generated by the TMD.

The inclusion of transverse momenta from the PB-TMD and the parton shower is performed using the \cascade\ event generator~\cite{Baranov:2021uol}. The factorization scale $\mu$ choice was chosen according to the performance of the matched PB-TMD + \powheg\ calculation on the differential jet rate $d_{01}$, which represents the square of the energy scale at which a 0--jet event is resolved as an 1--jet event. A smoother $d_{01}$ implies a better matching between the PB-TMD shower contribution to the 0--parton calculation and the real emission contribution. The choice $\mu = \sum_{i}\sqrt{m_i^2+p_{t,i}^2}$ for 1--parton events and $\mu = \sqrt{\hat{s}}$ for 0--parton events (where $\hat{s}$ is the partonic center-of-mass energy) showed the smoothest $d_{01}$ distribution. This option has been added to the \cascade\ event generator.

\begin{figure}
\begin{center} 
\includegraphics[width=0.50\textwidth]{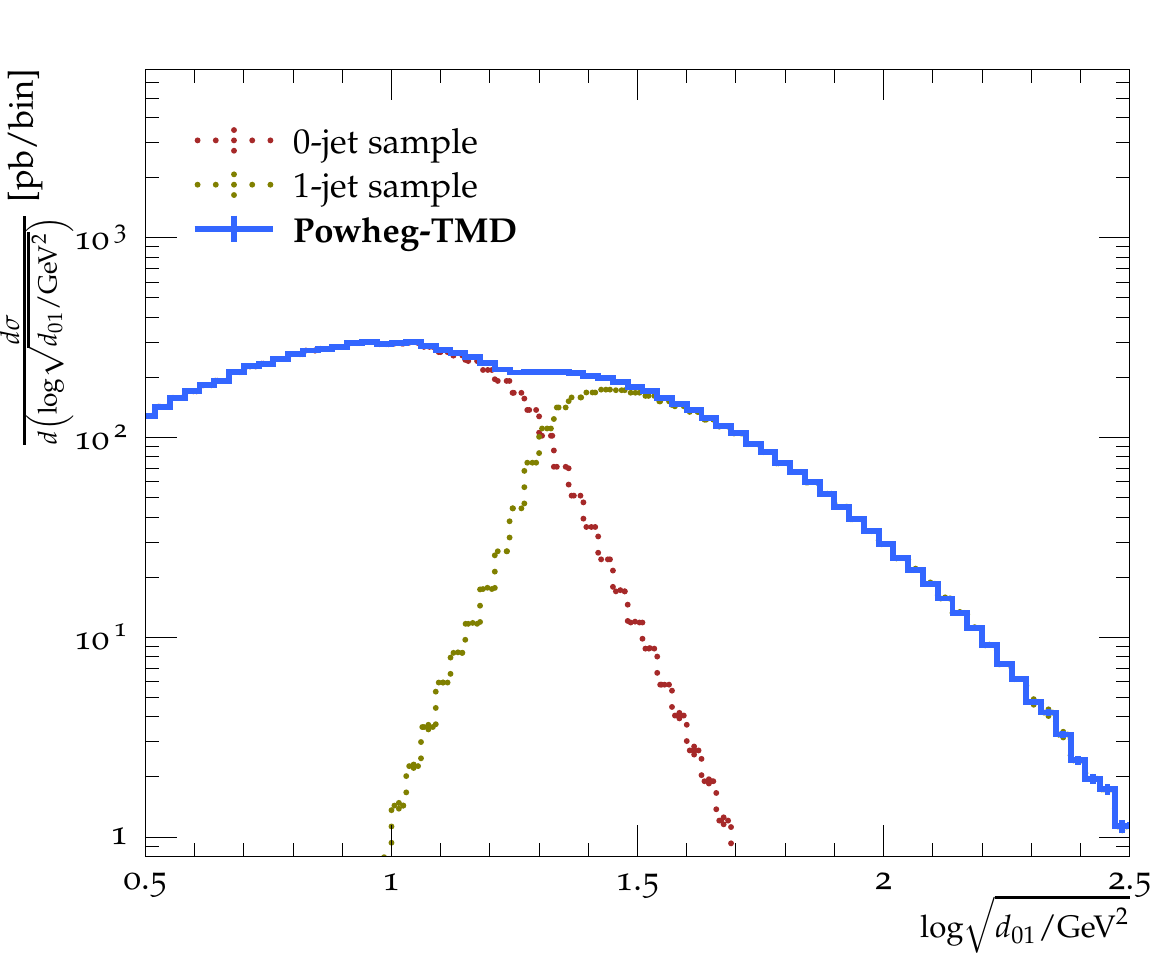}
\includegraphics[width=0.45\textwidth]{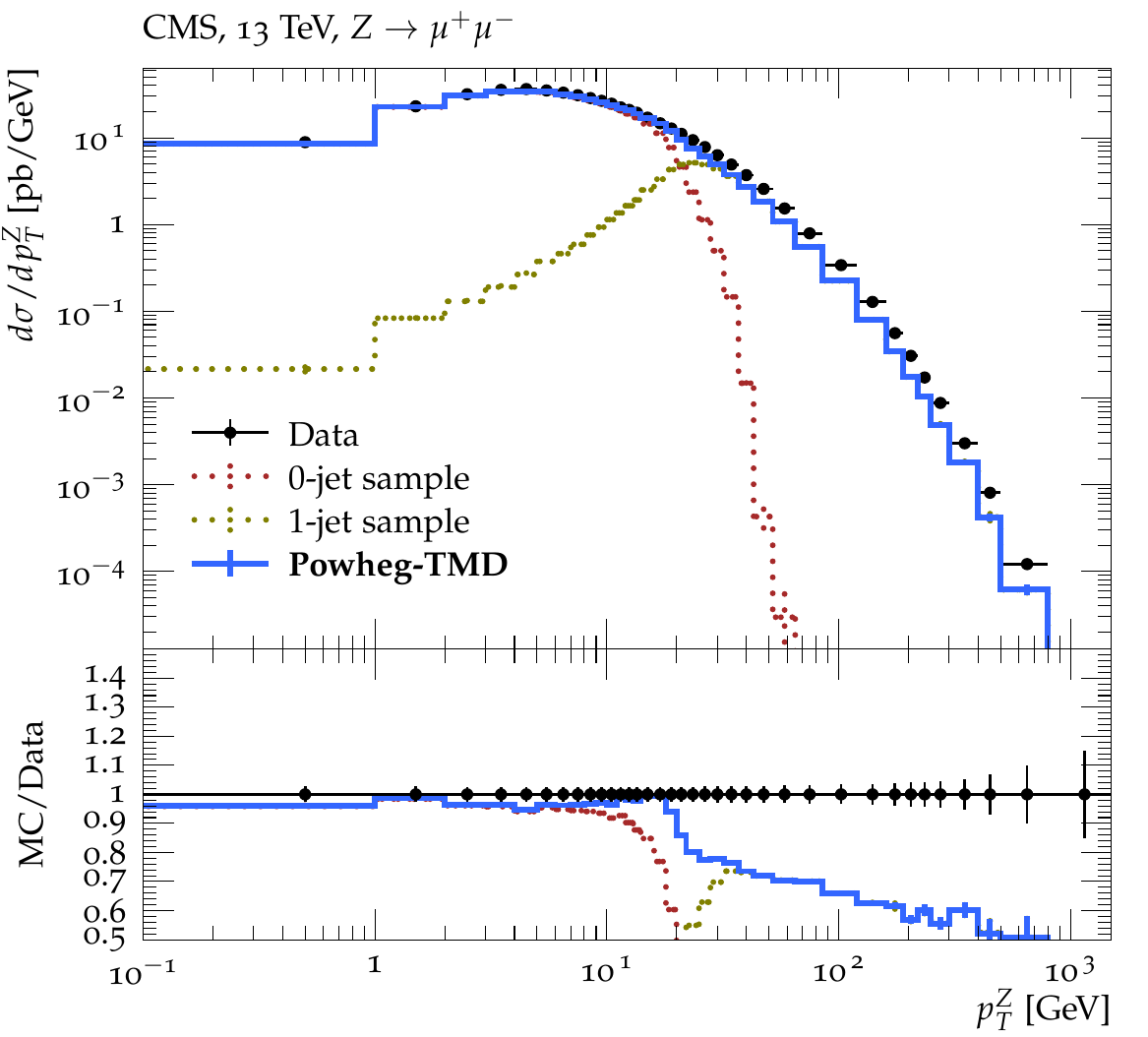}
  \caption{\small  (Left) The $d_{01}$ spectrum representing the energy-square scale at which an 0--jet event is resolved as an 1--jet event. The dotted curves represent the contributions of the single-multiplicity samples. (Right) DY \pt\ spectrum as obtained from a matched PB-TMD + \powheg\ computation compared to the measurement of the DY \pt\ spectrum by the CMS experiment at the center-of-mass energy of 13 TeV~\protect\cite{CMS:2019raw}.}
\label{fig3}
\end{center}
\end{figure}

In Fig.~\ref{fig3} the described matching procedure is applied to DY production. In Fig.~\ref{fig3} (left) the $d_{01}$ distribution is shown together with the contributions from the 0-- and 1--jet calculations matched to the PB-TMD parton shower. In Fig.~\ref{fig3} (right) the computation of the DY \pt\ spectrum is depicted.     

As observed in Fig.~\ref{fig3}, a good matching of the PB-TMD with \powheg\ NLO calculation of DY production is achieved, resulting in a good description of the DY \pt\ spectrum at low and intermediate \pt\ . The results are in agreement with the previous results obtained using \mcatnlo\ matching~\cite{BermudezMartinez:2019anj,BermudezMartinez:2020tys}. The difference observed between the data and the calculation at high \pt\ is due to missing higher orders. 

We should point out that further refinements of the method are under investigation. Among these is to avoid the posibility of having parton shower emissions that are are harder than the \powheg\ real emission even though the resulting transverse momentum from the TMD is not. This effect is due to the angular ordering of the partonic emissions in the PB-TMD parton shower.  

\section{Conclusion}

The DESY summer student program ran exclusively online in its 2021 edition. In a dedicated effort, the students worked on the matching of PB-TMDs determined from NLO fits to DIS data, and the resulting PB-TMD parton shower, to NLO calculation using the \powheg\ method. The computation has been compared to the DY \pt\ spectrum measured by the CMS collaboration at the center-of-mass energy of 13 TeV. A very good agreement at small and intermediate \pt\ was observed, which is also compatible with results obtained using the \mcatnlo\ method~\cite{BermudezMartinez:2019anj,BermudezMartinez:2020tys}. We should poin out that this performance is achived without further adjustment of parton shower parameters as opposed to collinear parton shower approaches. The new scale choice option for the matching has been included in the \cascade\ event generator.

Although a smooth $d_{01}$ distribution distribution has been observed, further refinement of the calculation might be needed in order to avoid remaining double counting of emissions, which has its origin in the angular ordering of the PB-TMD parton emissions.

\bibliographystyle{mystyle}

\begin{thebibliography}{99}
\bibitem{Drell:1970wh}
S.~Drell and T.-M. Yan,
Phys.Rev.Lett.{} {\bf 25},~316~(1970)\relax
\relax
\bibitem{Dokshitzer:1978yd}
Y.~L.~Dokshitzer, D.~Diakonov and S.~I.~Troian,
Phys. Lett. B \textbf{79} (1978), 269-272\relax
\relax
\bibitem{Parisi:1979se}
G.~Parisi and R.~Petronzio,
Nucl. Phys. B \textbf{154} (1979), 427-440\relax
\relax
\bibitem{Curci:1979am}
G.~Curci and M.~Greco,
Phys. Lett. B \textbf{92} (1980), 175-178\relax
\relax
\bibitem{Altarelli:1984kp}
G.~Altarelli, R.~K.~Ellis and G.~Martinelli,
Z. Phys. C \textbf{27} (1985), 617\relax
\relax
\bibitem{Collins:1984kg}
J.~C.~Collins, D.~E.~Soper and G.~F.~Sterman,
Nucl. Phys. B \textbf{250} (1985), 199-224\relax
\relax
\bibitem{Bizon:2018foh}
W.~Bizon, X.~Chen, A.~Gehrmann-De Ridder, T.~Gehrmann, N.~Glover, A.~Huss, P.~F.~Monni, E.~Re, L.~Rottoli and P.~Torrielli,
JHEP \textbf{12} (2018), 132
[arXiv:1805.05916]\relax
\relax
\bibitem{Bizon:2019zgf}
W.~Bizon, A.~Gehrmann-De Ridder, T.~Gehrmann, N.~Glover, A.~Huss, P.~F.~Monni, E.~Re, L.~Rottoli and D.~M.~Walker,
Eur. Phys. J. C \textbf{79} (2019) no.10, 868
[arXiv:1905.05171]\relax
\relax
\bibitem{Catani:2015vma}
S.~Catani, D.~de Florian, G.~Ferrera and M.~Grazzini,
JHEP \textbf{12} (2015), 047
[arXiv:1507.06937]\relax
\relax
\bibitem{Scimemi:2017etj}
I.~Scimemi and A.~Vladimirov,
Eur. Phys. J. C \textbf{78} (2018) no.2, 89
[arXiv:1706.01473]\relax
\relax
\bibitem{Bacchetta:2019tcu}
A.~Bacchetta, G.~Bozzi, M.~Lambertsen, F.~Piacenza, J.~Steiglechner and W.~Vogelsang,
Phys. Rev. D \textbf{100} (2019) no.1, 014018
[arXiv:1901.06916]\relax
\relax
\bibitem{Bacchetta:2018lna}
A.~Bacchetta, G.~Bozzi, M.~Radici, M.~Ritzmann and A.~Signori,
Phys. Lett. B \textbf{788} (2019), 542-545
[arXiv:1807.02101]\relax
\relax
\bibitem{Ladinsky:1993zn}
G.~A.~Ladinsky and C.~P.~Yuan,
Phys. Rev. D \textbf{50} (1994), R4239
[arXiv:hep-ph/9311341]\relax
\relax
\bibitem{Balazs:1997xd}
C.~Balazs and C.~P.~Yuan,
Phys. Rev. D \textbf{56} (1997), 5558-5583
[arXiv:hep-ph/9704258]\relax
\relax
\bibitem{Landry:2002ix}
F.~Landry, R.~Brock, P.~M.~Nadolsky and C.~P.~Yuan,
Phys. Rev. D \textbf{67} (2003), 073016
[arXiv:hep-ph/0212159]\relax
\relax
\bibitem{Nadolsky}
P.~Nadolsky et al.,
http://hep.pa.msu.edu/resum/
\relax
\bibitem{Alioli:2015toa}
S.~Alioli, C.~W.~Bauer, C.~Berggren, F.~J.~Tackmann and J.~R.~Walsh,
Phys. Rev. D \textbf{92} (2015) no.9, 094020
[arXiv:1508.01475]\relax
\relax
\bibitem{Bozzi:2019vnl}
G.~Bozzi and A.~Signori,
Adv. High Energy Phys. \textbf{2019} (2019), 2526897
[arXiv:1901.01162]\relax
\relax
\bibitem{Baranov:2014ewa}
S.~P.~Baranov, A.~V.~Lipatov and N.~P.~Zotov,
Phys. Rev. D \textbf{89} (2014) no.9, 094025
[arXiv:1402.5496]\relax
\relax
\bibitem{Sjostrand:2014zea}
T.~Sj\"ostrand, S.~Ask, J.~R.~Christiansen, R.~Corke, N.~Desai, P.~Ilten, S.~Mrenna, S.~Prestel, C.~O.~Rasmussen and P.~Z.~Skands,
Comput. Phys. Commun. \textbf{191} (2015), 159-177
[arXiv:1410.3012]\relax
\relax
\bibitem{Bellm:2015jjp}
J.~Bellm, S.~Gieseke, D.~Grellscheid, S.~Pl\"atzer, M.~Rauch, C.~Reuschle, P.~Richardson, P.~Schichtel, M.~H.~Seymour and A.~Si\'odmok, \textit{et al.}
Eur. Phys. J. C \textbf{76} (2016) no.4, 196
[arXiv:1512.01178]\relax
\relax
\bibitem{Bahr:2008pv}
M.~Bahr, S.~Gieseke, M.~A.~Gigg, D.~Grellscheid, K.~Hamilton, O.~Latunde-Dada, S.~Platzer, P.~Richardson, M.~H.~Seymour and A.~Sherstnev, \textit{et al.}
Eur. Phys. J. C \textbf{58} (2008), 639-707
[arXiv:0803.0883]\relax
\relax
\bibitem{Gleisberg:2008ta}
T.~Gleisberg, S.~Hoeche, F.~Krauss, M.~Schonherr, S.~Schumann, F.~Siegert and J.~Winter,
JHEP \textbf{02} (2009), 007
[arXiv:0811.4622]\relax
\relax
\bibitem{Frixione:2003ei}
S.~Frixione, P.~Nason and B.~R.~Webber,
JHEP \textbf{08} (2003), 007
[arXiv:hep-ph/0305252]\relax
\relax
\bibitem{Frixione:2002ik}
S.~Frixione and B.~R.~Webber,
JHEP \textbf{06} (2002), 029
[arXiv:hep-ph/0204244]\relax
\relax
\bibitem{Frixione:2007vw}
S.~Frixione, P.~Nason and C.~Oleari,
JHEP \textbf{11} (2007), 070
[arXiv:0709.2092]\relax
\relax
\bibitem{Nason:2012pr}
P.~Nason and B.~Webber,
Ann. Rev. Nucl. Part. Sci. \textbf{62} (2012), 187-213
[arXiv:1202.1251]\relax
\relax
\bibitem{Alwall:2014hca}
J.~Alwall, R.~Frederix, S.~Frixione, V.~Hirschi, F.~Maltoni, O.~Mattelaer, H.~S.~Shao, T.~Stelzer, P.~Torrielli and M.~Zaro,
JHEP \textbf{07} (2014), 079
[arXiv:1405.0301]\relax
\relax
\bibitem{Frederix:2015eii}
R.~Frederix, S.~Frixione, A.~Papaefstathiou, S.~Prestel and P.~Torrielli,
JHEP \textbf{02} (2016), 131
[arXiv:1511.00847]\relax
\relax
\bibitem{BermudezMartinez:2019anj}
A.~Bermudez Martinez, P.~Connor, D.~Dominguez Damiani, L.~I.~Estevez Banos, F.~Hautmann, H.~Jung, J.~Lidrych, M.~Schmitz, S.~Taheri Monfared and Q.~Wang, \textit{et al.}
Phys. Rev. D \textbf{100} (2019) no.7, 074027
[arXiv:1906.00919]\relax
\relax
\bibitem{BermudezMartinez:2020tys}
A.~Bermudez Martinez, P.~L.~S.~Connor, D.~Dominguez Damiani, L.~I.~Estevez Banos, F.~Hautmann, H.~Jung, J.~Lidrych, A.~Lelek, M.~Mendizabal and M.~Schmitz, \textit{et al.}
Eur. Phys. J. C \textbf{80} (2020) no.7, 598
[arXiv:2001.06488]\relax
\relax
\bibitem{Hautmann:2017xtx}
F.~Hautmann, H.~Jung, A.~Lelek, V.~Radescu and R.~Zlebcik,
Phys. Lett. B \textbf{772} (2017), 446-451
[arXiv:1704.01757]\relax
\relax
\bibitem{Hautmann:2017fcj}
F.~Hautmann, H.~Jung, A.~Lelek, V.~Radescu and R.~Zlebcik,
JHEP \textbf{01} (2018), 070
[arXiv:1708.03279]\relax
\relax
\bibitem{CMS:2019raw}
A.~M.~Sirunyan \textit{et al.} [CMS],
JHEP \textbf{12} (2019), 061
[arXiv:1909.04133]\relax
\relax
\bibitem{ATLAS:2015iiu}
G.~Aad \textit{et al.} [ATLAS],
Eur. Phys. J. C \textbf{76} (2016) no.5, 291
[arXiv:1512.02192]\relax
\relax
\bibitem{NuSea:2003qoe}
J.~C.~Webb \textit{et al.} [NuSea],
[arXiv:hep-ex/0302019]\relax
\relax
\bibitem{Webb:2003bj}
J.~C.~Webb,
[arXiv:hep-ex/0301031]\relax
\relax
\bibitem{Moreno:1990sf}
G.~Moreno, C.~N.~Brown, W.~E.~Cooper, D.~Finley, Y.~B.~Hsiung, A.~M.~Jonckheere, H.~Jostlein, D.~M.~Kaplan, L.~M.~Lederman and Y.~Hemmi, \textit{et al.}
Phys. Rev. D \textbf{43} (1991), 2815-2836
\relax
\bibitem{PHENIX:2018dwt}
C.~Aidala \textit{et al.} [PHENIX],
Phys. Rev. D \textbf{99} (2019) no.7, 072003
[arXiv:1805.02448]\relax
\relax
\bibitem{Martinez:2021chk}
A.~B.~Martinez, F.~Hautmann and M.~L.~Mangano,
Phys. Lett. B \textbf{822} (2021), 136700
[arXiv:2107.01224]\relax
\relax
\bibitem{BermudezMartinez:2018fsv}
A.~Bermudez Martinez, P.~Connor, H.~Jung, A.~Lelek, R.~\v{Z}leb\v{c}\'\i{}k, F.~Hautmann and V.~Radescu,
Phys. Rev. D \textbf{99} (2019) no.7, 074008
[arXiv:1804.11152]\relax
\relax
\bibitem{Alioli:2010xd}
S.~Alioli, P.~Nason, C.~Oleari and E.~Re,
JHEP \textbf{06} (2010), 043
[arXiv:1002.2581]\relax
\relax
\bibitem{Nason:2004rx}
P.~Nason,
JHEP \textbf{11} (2004), 040
[arXiv:hep-ph/0409146]\relax
\relax
\bibitem{Alwall:2006yp}
J.~Alwall, A.~Ballestrero, P.~Bartalini, S.~Belov, E.~Boos, A.~Buckley, J.~M.~Butterworth, L.~Dudko, S.~Frixione and L.~Garren, \textit{et al.}
Comput. Phys. Commun. \textbf{176} (2007), 300-304
[arXiv:hep-ph/0609017]\relax
\relax
\bibitem{Baranov:2021uol}
S.~Baranov, A.~Bermudez Martinez, L.~I.~Estevez Banos, F.~Guzman, F.~Hautmann, H.~Jung, A.~Lelek, J.~Lidrych, A.~Lipatov and M.~Malyshev, \textit{et al.}
Eur. Phys. J. C \textbf{81} (2021) no.5, 425
[arXiv:2101.10221]\relax
\relax
\end{thebibliography}
\raggedright

\end{document}